\documentclass{tlp}

\usepackage{aopmath}

\newtheorem{definition}{Definition}

\begin{document}

\title[Three-valued semantics]
      {A three-valued semantics for logic programmers}

\author[L. Naish]
{LEE NAISH\\
Department of Computer Science and Software Engineering \\
University of Melbourne \\
Melbourne 3010 \\
Australia\\
\email{lee@cs.mu.oz.au}}

\submitted{15 June 2004}
\revised{30 November 2004}
\accepted{30 June 2005}

\maketitle

\begin{abstract}

This paper describes a simpler way for programmers to reason about the
correctness of their code.  The study of semantics of logic programs
has shown strong links between the model theoretic semantics (truth
and falsity of atoms in the programmer's interpretation of a program),
procedural semantics (for example, SLD resolution) and fixpoint semantics
(which is useful for program analysis and alternative execution
mechanisms).  Most of this work assumes that intended interpretations
are two-valued: a ground atom is true (and should succeed according to
the procedural semantics) or false (and should not succeed).  In reality,
intended interpretations are less precise.  Programmers consider that some
atoms ``should not occur'' or are ``ill-typed'' or ``inadmissible''.
Programmers don't know and don't care whether such atoms succeed.
In this paper we propose a three-valued semantics for (essentially) pure
Prolog programs with (ground) negation as failure which reflects this.
The semantics of Fitting is similar but only associates the third
truth value with non-termination.  We provide tools to reason about
correctness of programs without the need for unnatural precision or undue
restrictions on programming style.  As well as theoretical results, we
provide a programmer-oriented synopsis.  This work has come out of work
on declarative debugging, where it has been recognised that inadmissible
calls are important.
This paper has been accepted to appear in Theory and Practice of Logic
Programming.
\end{abstract}
\begin{keywords}
Models, immediate consequence operator, SLDNF resolution, negation,
verification, declarative debugging, inadmissibility
\end{keywords}

\section{Introduction}

When programming, we often give too little attention to the meaning
of our code.  For these sins of sloth and pride (``I can't be bothered
re-checking everything --- I'm sure I got it right'') we are forced to
do penance, in the form of debugging.  During the repetitive tedium
we contemplate the relationships between our code, its behaviour and
our desires: the domain of programming language semantics.  Thus (in
retrospect) it is natural for work on debugging to lead to work
on semantics.  All too often, work on semantics seems to bear little
relation to any stage of the software life cycle.  Here our aim is to
provide theoretical support which can allow programmers to reason about
the correctness of their code more easily.

The starting point for our three-valued approach to semantics was
declarative debugging of logic programs.  The conventional view of
the semantics of logic programs \cite{Llo84} and declarative debugging
\cite{Sha83} \cite{Llo87} is as follows.  Every ground atom is either true
or false in the intended interpretation of the program.  True atoms should
succeed and false atoms should fail.  To isolate a bug a declarative
debugger compares the intended interpretation with the behaviour of a
program in a particular instance which terminates and which produces
unexpected results.  It has been noted by various researchers that
this strict division into true and false does not correspond to the
intentions of typical programmers.  There are some atoms encountered
during debugging which simply should not occur in correct programs (for
example, ``ill-typed'' atoms).  Whether they succeed or fail is not the
issue.  A third truth value, \emph{inadmissible}, has been introduced
into declarative debugging for such atoms \cite{Per86} \cite{PerCal88}
\cite{ddscheme3}.  This is a recognition that, in practice, \emph{intended
interpretations are three-valued} rather than two-valued.  Similarly,
most work on formal specifications allows for the behaviour of a module
to be unspecified if \emph{preconditions} are violated.  The aim of this
paper is to reconstruct the semantics of logic programs with this in mind.


There are many reasons for studying programming language semantics.
One is pure philosophy --- knowing more about a language for the
sake of knowing.  A formal semantics is also a useful guide for an
implementer and allows programmers to write portable code which has
predictable behaviour.  It can also be used for optimisation and
analysis of programs, to help make implementations more efficient as
well as correct.  Semantics can also provide useful ways for programmers
to think about their code, and can be the basis for program development
environments including debuggers.  This is the main focus for our work ---
we provide a semantics for \emph{logic programmers} (though implementers
and philosophers may also be interested).  We aim to allow programmers to
reason about the partial correctness of programs as easily as possible.
Model theory is particularly attractive due to its simplicity --- partial
correctness can be ensured if the intended interpretation is a model
of each clause in the program.  It is a useful guide for constructing
programs, an excellent tool for verification of programs and enables
declarative debugging.  Finally, all the considerations above can be
used in the design of programming languages.

There is (very) extensive work on negation in logic programming
(see \cite{apt94logic}) including some three-valued approaches
\cite{Fitting85}\cite{Kun87}.  The work tends to be directed at
solving mathematical and computational problems in order to make the
declarative and procedural semantics as close as possible, and there is
a general desire to be able to assign a meaning to all possible programs.
In our view, both these goals have questionable benefit for programmers.
The declarative semantics should support a notion of inadmissibility which
is not tied down to any particular program behaviour, thus leading to
an inevitable gap between declarative and procedural readings.  A key
question for programmers is whether a program is correct according
to their intended interpretation.  If the answer is no, one possible
solution is to change the intended interpretation --- ``its not a bug,
its a feature'' (having a semantics for every program is useful here).
However, the more common solution is to change the program!

Because this paper is partly concerned with how programmers think,
some qualification is in order.  I am not a psychologist and the ideas
here were not the result of a proper study of programmers.  In fact, I
believe most Prolog programmers think little or nothing of declarative
semantics (though perhaps with a better semantics this will change).
The ideas came mostly from introspection and are a refinement of the
work I have done over many years on reasoning about logic programs.
The intended meanings of programs in this paper are discussed and I
know the intended meanings because I wrote the programs.  I hope to
convince the reader that other programmers think in a similar way, or at
least this is a useful way to think about programs.  I believe my use
of declarative semantics has contributed to my logic
programming abilities.

The rest of the paper is structured as follows.  We first give a synopsis
for programmers: a concise non-technical description of a verification
method which arises out of our technical results.  The next section gives
more detailed motivation for our work, discussing how programmers intend
code to behave, how it actually behaves, reasoning about (in)correctness
and formal semantics.  It briefly reviews previous work on the semantics
of logic programs and declarative debugging (some knowledge of these areas
would help the reader).  We then present our three-valued semantics,
first for definite clause programs, then for programs with negation.
We define operational, model theoretic and fixpoint semantics and
establish various relationships between them.  Related work on semantics
is discussed in these sections.  Finally, we present examples of our
program verification method, comparing it with other work, then conclude.

\section{Synopsis for programmers}

The theoretical results in this paper show you can establish that a program
\begin{enumerate}
\item returns no wrong answers, and
\item misses no answers in all solutions computations which terminate
normally,
\end{enumerate}
using the following verification method:
\begin{enumerate}
\item
You must decide whether each ground atomic goal is \emph{true} (should
succeed), \emph{false} (should fail) or \emph{inadmissible} (should not
occur; any behaviour is acceptable).
\item
For each true atom there must be a ground matching clause instance
with a true body (all conjuncts are true).
\item
For each false atom, all ground matching clause instances
must have false bodies (at least one conjunct is false).
\end{enumerate}

Note that separate reasoning must be used to show the program terminates
normally (answers may be missed due to non-termination or runtime errors).
Also, there are several assumptions about the program/system:
\begin{enumerate}
\item  Only ``pure'' Prolog is used --- there are no uses of \emph{cut},
\texttt{var/1}, \emph{et cetera}.
\item Negated calls are ground when they are selected (this is checked at
runtime or compile time in some logic programming systems; others provide
no support for checking).
\item The execution doesn't violate the ``occurs check'' (again, some
systems provide support for this but most do not --- occurs check
violations are very rare).
\end{enumerate}

\section{Motivation and background}
\subsection{Conventional definite clause semantics}

The conventional approach to the semantics of logic programs, described in
\cite{Llo84}, includes model theory, fixpoint theory and an operational
semantics.  Here we just discuss the theory as it applies to successful
computations using definite clauses.  That is, we do not (yet) deal with
failure or negation, which lead to significant additional complication
and various conflicting proposed semantics.

The model theory is two-valued --- each (ground) atom in the Herbrand
base is associated with the truth value \emph{true} or \emph{false}.
Typically the theory represents an interpretation as a single set of
atoms --- the atoms which are true.  The classical truth tables for
conjunction and implication are used.  The intended interpretation of
a (correct) program is assumed to be a model (every clause instance
is true according to the interpretation and truth tables) and this
implies soundness of computations (see below).  This is the great
advantage of the model-theoretic semantics: a program can be viewed
completely declaratively and the correctness of each clause can be
verified statically and in isolation.  It can also be used as the basis
for declarative debugging.  The intersection of two models is a model,
hence a least model exists which is the intersection of all models and
is the least Herbrand model.  This is the set of logical consequences
of the program.

The fixpoint semantics of a program $P$ are based on the
immediate consequence
operator $T_P$, which maps a set of ground atoms $M$ to the set of ground
atoms which can be proven from $M$ by applying a single program clause:
\medskip\\
$T_P(M) = \{H | H \leftarrow B_1, B_2, \ldots B_N$ is a
ground clause instance \\
\hspace*{2cm} and $\{B_1, B_2, \ldots B_N\} \subseteq M\}$
\medskip\\
$T_P$ is monotonic, and applying $T_P$ $n$ times starting with the
empty set ($T_P \uparrow n$) is of interest.  It gives a ``bottom up''
semantics where initially we assume nothing and iteratively prove that a
growing set of atoms are true.  The least fixpoint of
$T_P$ is $T_P \uparrow \omega$
(= $\textrm{lub}(T_P \uparrow n), n = 1,2, \ldots$).
A set of atoms $M$ is a model of $P$ if and only if $T_P(M) \subseteq M$
and the least fixpoint is the
least model.  The fixpoint semantics are particularly useful for
program analysis and have also been used as the basis for bottom up
operational semantics, especially for logic databases.

The operational model is SLD resolution.  This gives a top down semantics
where atoms are proved by recursively proving the bodies of matching
clauses.  The set of ground atoms which have successful SLD derivations
(which is the set of ground instances of computed answers) is called
the success set, $SS$, and is independent of the computation rule (the
order in which atoms are selected).  The success set is the same as the
least model and the least fixpoint.  Thus (assuming that the intended
interpretation is a model) the computed answers of a program are true
in the intended interpretation.

\subsection{Intended interpretations are not models!}

\begin{figure}
\figrule
\begin{verbatim}
merge([], Bs, Bs).
merge(A.As, [], A.As).
merge(A.As, B.Bs, A.Cs) :- A =< B, merge(As, B.Bs, Cs).
merge(A.As, B.Bs, B.Cs) :- A > B, merge(A.As, Bs, Cs).
\end{verbatim}
\caption{Merging sorted lists of numbers}
\figrule
\label{fig_merge}
\end{figure}

The major flaw in this otherwise beautiful picture is that, in practice,
the intended interpretation is generally \emph{not} a model.
%
Consider the following informal predicate description.
\begin{verbatim}
merge(As, Bs, Cs): Cs is the sorted list of numbers
which is the multiset union of sorted lists As and Bs
\end{verbatim}
This would make a reasonable comment attached to some code (for example,
Figure\ \thefigure) or a specification for a programmer (for example,
an exercise in an introductory Prolog course).  In addition, it might
be helpful to say that \texttt{As} and \texttt{Bs} are intended to be
input, though there is no notion of input and output in the conventional
declarative semantics.  Programmers have little difficulty with such
descriptions of the intended behaviour of predicates.  It could be
surmised that such descriptions correspond to intended interpretations of
the model theory and that the primary job of a programmer is to construct
programs for which such interpretations are models.  However, critical
examination reveals this to be false, even for very simple programs.

First, the intended interpretation being a model only guarantees soundness
(partial correctness or no wrong answers).  In reality, some form of
completeness is also necessary: if every predicate always fails then
the intended interpretation will be a model but the program will not be
very useful!  Without considering negation, the most obvious way to avoid
missing answers is for the intended interpretation to be the minimum model
(which is the same as the success set).  We discuss another possibility
in section\ \ref{sec_bg_neg}.

Second, consider the first clause for \texttt{merge/3} (the same as
the base case for the well known \texttt{append/3} program).  It can
succeed with non-lists, and precisely which calls to \texttt{merge/3}
containing non-lists succeed is a subtle property of the code and not
dealt with at all in the predicate description (from the description
we might assume that no such calls should succeed).  The description is
symmetric with respect to the first two arguments but the code is not,
and the set of successful calls with non-lists is not.  If the code was
changed so the first two arguments of each \texttt{merge/3} atom were
swapped, we would have a different definition with a different success
set and minimum model.  Similarly, we could swap these two arguments
in just recursive calls --- either or both of them.  Which (if any)
of these is correct according to the (single) intended interpretation?

\subsection{Types, assertions and preconditions}

Strong static typing such as that used in G\"{o}del \cite{goedel} and
Mercury \cite{mercury} overcome problems with non-lists --- we don't
have to consider if \texttt{merge/3} behaves correctly with non-lists
because it can't be called with or return non-lists.  However, there
are similar problems with unsorted lists:  \texttt{merge/3} can succeed
with unsorted lists, it is asymmetric and there are many different
definitions with different behaviours for unsorted lists.  When code is
written there are nearly always implicit assumptions about the context
in which it is called.  Often the assumptions about the calling context
concern what are normally called types (for example, lists), though
they can also be more complex and impossible to check statically.
There is certainly some merit in having such assumptions documented by
programmers declaring (possibly more complex) ``types'', ``assertions''
or ``preconditions''; this was our approach in \cite{naish:90}.  As
well as improving code readability, it can help with finding bugs
because the extra information, which is redundant in correct programs,
may be inconsistent with the executable part of the program.  The
Mercury compiler requires declarations concerning types, modes and
determinism, and uses this information to produce more efficient code.
Optional declarations are also used to increase efficiency in systems
such as Ciao \cite{DBLP:conf/discipl/PueblaBH00}.  A more expressive
language for defining preconditions could help the Mercury compiler
recognise determinism in more cases, for example, or allow certain
transformations which would not be correct in general but are correct
if the precondition holds.

However, there are several disadvantages and limitations of a semantics
which relies on declared preconditions.  First, as well as the language
used to define \texttt{merge/3}, we have a language (with possibly a
different syntax and its own semantics) for defining sorted lists and
(typically) a declaration language relating the two.  The semantics, and
preferably debugging environments, are more complex and must deal with
unintended cases such as preconditions which are always false.  Second,
it is hard to convince programmers to add extra declarations to their
programs when they are unnecessary for the procedural semantics.  Third,
even willing programmers are unable to document all their assumptions
using formal languages.  For example, code for many declarative debuggers
(discussed in the next section) assume that certain predicates are
called with (representations of) atoms which have \emph{finite} SLD
trees or refutations and even assume the atoms have particular truth
values in the intended interpretation \cite{naish-92}.  Understanding the
preconditions allows us to reason about the soundness and completeness
of the debuggers.  However, defining them formally would need us to
circumvent the undecidability of the halting problem and formalise the
intentions of certain programmers in the future!

Our approach to the semantics of a program is based on the view that
the behaviours we care about are a subset of its possible behaviours.
The subset may be described in some way but is not necessarily codified
precisely (and it may not even be possible to do so).  For example, we
could use type and mode declarations to document the fact we care only
about the behaviour of \texttt{merge/3} when the first two arguments are
instantiated to lists of numbers (and this information could be used for
program analysis and optimisation).  The semantics we propose is thus
compatible with various forms of preconditions.  However, it also leaves
open the possibility that there are further undocumented restrictions
on what we care about, such as the sortedness of the lists.

\subsection{Three-valued declarative debugging}

A general scheme for declarative debugging using three truth values
is described in \cite{ddscheme3}; it is based on a more traditional
two-valued scheme \cite{ddscheme}.  Computations are represented as trees
and debuggers search the tree for a \emph{buggy} node.  For diagnosing
wrong answers in definite clause programs, the tree is a proof tree
(see \cite{Llo84}): each node contains an atom $A$ which was proved
in the computation and the children of the node contain the atoms in
the body of the clause instance which was used at the top level of
the proof of $A$.  Each node has a truth value associated with it:
\emph{correct}, if the atom is true in the intended interpretation,
\emph{erroneous}, if the atom is false, or \emph{inadmissible}, the
third truth value\footnote{For simplicity we ignore non-ground atoms
here.}.  The truth value is determined by an ``oracle''.  The user is
asked questions and typically these are stored in a database to avoid
repeated questions.  The user may also be able to supply more general
assertions or even runnable specifications \cite{Drabent-88}.

If a node is erroneous but all its children are correct, it corresponds
to an \emph{incorrect clause instance}: the body is true but the head
is false.  This class of bugs was identified in the first work on
declarative debugging \cite{Sha83}.  Another class of bugs, related to
inadmissibility, was identified in \cite{Per86} and formalised more in
\cite{ddscheme3}: nodes which are erroneous with no erroneous children but
at least one inadmissible child.  In a top-down execution this corresponds
to a clause instance which causes a transition from admissible atoms to
inadmissible atoms.  It allows an inadmissible atom to be used in the
(dubious) ``proof'' of a false atom.  For such a bug to be manifest,
the inadmissible call must succeed (if the inadmissible call fails the
top level false atom would also fail, so there would be no bug symptom
to diagnose).  However, the diagnosis algorithm does not consider how
or why the inadmissible call succeeds.  The fact it succeeds is not
considered an error --- the error is that it is called at all.

Several different instances of the three-valued declarative debugging
scheme were identified, using different definitions of inadmissibility.
Although our previous work on types and debugging provides the
intuition for this paper, we do not rely on any particular definition
of inadmissibility here.  We simply assume that the programmer has
some notion of (in)admissibility for ground atoms.  If inadmissibility
is identified with ill-typedness (as suggested in \cite{Per86}) then
the second class of bugs correspond to a form of type error discussed
in \cite{naish:90}.  However, this definition of inadmissibility does
not lead to ideal behaviour of debuggers \cite{ddscheme3} and does not
quite capture the intentions of programmers.

As well as considering types, programmers consider modes.  A successful
call to \verb@merge/3@ in which just the last argument is not a sorted
list is very different from a successful call where the second argument
is not a sorted list.  For \texttt{merge([2,3], [1,2], [2,1,2,3])} we
would consider that the output is wrong and there is a bug in \verb@merge/3@
whereas for \texttt{merge([2,3], [2,1], [2,1,2,3])} we would consider that
the input is wrong and the bug is elsewhere.  A more natural definition
of inadmissibility for ground atoms is that the ``input'' arguments of
the atom are ill-typed (or violate some condition).  In \cite{ddscheme3}
this definition of inadmissibility is related to a declarative view of
modes \cite{modes}, which gives a more technical definition that captures
the intuitive idea of input arguments being ill-typed.


\subsection{Missing answers and negation}
\label{sec_bg_neg}

As well as finding wrong answers, Prolog programs can miss correct
answers.  Missing answers can be diagnosed with the same algorithm
but a different kind of tree.   Tree nodes can contain calls together
with sets of answers returned and the children of a node can be all
the calls in the bodies of matching clauses.  A node is correct if all
correct answers are returned and can be considered inadmissible if the
input arguments violate some condition.  Diagnosis of both wrong and
missing answers can be done using a combination of both kinds of trees.
Such a combined tree can be used to diagnose bugs in program containing
negation --- ``\texttt{not p}'' returns a wrong answer if \texttt{p}
misses an answer and vice versa.

Treating inadmissible atoms as being true, though somewhat
counter-intuitive, results in accurate diagnoses using two-valued
declarative debugging if we restrict attention to a single wrong answer
diagnosis in a definite clause program.  This corresponds to saying
inadmissible atoms are true in the intended interpretation.  However, we
often repeatedly diagnose a bug, modify the program and re-test it until
the intended interpretation is a model.  If the oracle retains information
about the intended interpretation during this process, as we would expect,
or we are interested in diagnosing missing answers or negation is used, a
single two-valued intended interpretation can lead to incorrect diagnosis.


For example, suppose \texttt{merge([2,3], [2,1], X)} returns \texttt{X =
[2,1,2,3]} as the only answer and the user says it is true.  The user
must not say that \texttt{merge([2,3], [2,1], [2,2,1,3])} is true during
missing answer diagnosis or a bug would be incorrectly diagnosed in
\texttt{merge/3}.  In a later version of the program \texttt{merge/3}
may have been modified (to fix a real or imagined bug or make it more
general or more efficient), so \texttt{merge([2,3], [2,1], X)} returns
\texttt{X = [2,2,1,3]} as the only answer.  Missing answer diagnosis
of \texttt{merge([2,3], [2,1], [2,1,2,3])} would then be incorrect
(since the atom fails but has been previously declared to be true).
Similarly, if the oracle knew that \texttt{merge/3} was intended to be
a function in this mode (from a user assertion or a declaration in the
program, such as ``det'' in Mercury or Ciao), wrong answer diagnosis
of \texttt{merge([2,3], [2,1], [2,2,1,3])} would be incorrect (this is
another use of negation, in the oracle rather than the program).

There are several possible solutions to these problems, but at their
core is a three-valued interpretation.  One solution is for the debugger
to be more procedural, saying whether an atom succeeded or failed and
asking if that behaviour is correct.  Inadmissible atoms are precisely
those for which both success and failure are considered correct.
Another solution is to use two separate two-valued interpretations (an
upper and lower bound on what is expected to succeed) for diagnosis of
wrong and missing answers, respectively.  The lower bound should be a
subset of the least model but (if it is not empty) means correct programs
cannot simply fail.  Inadmissible atoms are those with differing truth
values in these two interpretations.  The verification method proposed
in \cite{drabentTPLP}, which we discuss in more detail in Section
\ref{sec_verif}, essentially uses this approach.  The approach of
\cite{naish:90} which uses two separate programs, one with additional
``type'' (admissibility) checks, is also similar.  Atoms which succeed in
the original program and fail in the augmented program are inadmissible,
though there may also be other inadmissible atoms which fail in both
programs (so there is a procedural element to this approach which we
avoid here).

%
%
%
%
%
%
%
%
%
%
%
\begin{figure}
\figrule
\begin{tabular}{ll}
\texttt{even(N) :- e4(N). \% or e1/e2/e3      }&\texttt{ odd(N) :- o2(N).  \% or o1/o3/o4}
\smallskip\\
\texttt{e1(0).                                  }&\texttt{ o1(s(0)).}\\
\texttt{e1(s(s(N))) :- e1(N).                   }&\texttt{ o1(s(s(N))) :- o1(N).}
\smallskip\\
\texttt{e2(0).                                  }&\texttt{ o2(s(N)) :- even(N).}\\
\texttt{e2(s(N)) :- odd(N).                     }&\texttt{}
\smallskip\\
\texttt{e3(0).                                  }&\texttt{ o3(s(N)) :- not o3(N).}\\
\texttt{e3(s(N)) :- not e3(N).                  }&\texttt{}
\smallskip\\
\texttt{e4(N) :- not odd(N).                    }&\texttt{ o4(N) :- not even(N).}\\
\end{tabular}
\caption{Various definitions of \texttt{even/1} and \texttt{odd/1}}
\figrule
\end{figure}

Figure\ \thefigure\  shows some of the multitude of different ways
\texttt{even/1} and \texttt{odd/1} can be defined using ``successor''
notation for numbers.  Some of the 16 possible combinations (such as
the one shown, using \texttt{e4} and \texttt{o2}) rely on negation.
A semantics which allows us to easily check the partial correctness of
these programs is desirable.  Different version have different success
sets: in some versions various inadmissible atoms succeed (for example,
\texttt{even(s(s([])))}) and for one version nothing succeeds (everything
loops).  However, these differences do not reflect different programmer
intentions and the programmer does not know or care which inadmissible
atoms succeed in the different versions.  Any semantics which varies
between the different versions makes it impossible for the programmer
to first decide on the intended meaning then have the freedom to code
any of the versions.  The semantics we provide in this paper allows
the same intended model for all versions of even and odd, respectively,
giving this freedom.

The freedom to write looping programs is not a good thing in itself.
This is a disadvantage of our semantics but is an unavoidable consequence
of using a simple semantics which deals only with partial correctness.
Programmers do need to consider termination (and efficiency), but we
believe separate tools are desirable for these purposes.  Separation of
concerns can help with program construction and is particularly useful
for debugging, where termination can be observed instead of proven
(or conjectured or hoped for).

\subsection{Programs naturally have more than one meaning}

Much of the work on semantics of logic programs, particularly with
negation, attempts to define \emph{the} (unique) meaning of a logic
program.  Authors argue that the meaning they define is more natural than
any other.  For definite programs, the consensus amongst these authors
is that the minimum model is \emph{the} meaning (the set of true atoms).
We have argued that this is typically not the case.  Furthermore, there is
no unique meaning, even for definite programs.  Here we consider possible
meanings of the \texttt{merge/3} program (Figure\ \ref{fig_merge}) again
(in Section\ \ref{sec_verif} we give further examples).  An alternative
description is as follows (a ``run'' being a maximal sorted sublist):
\begin{verbatim}
merge(As, Bs, Cs): Cs is an interleaving of the lists
of integers As and Bs and the number of runs in Cs is
the maximum number of runs in As and Bs
\end{verbatim}

This could be refined to define the interleaving more precisely.
Implementing this second specification (or verifying an implementation)
requires different reasoning from the first and the implementation is
likely to be used in different ways (a different version of merge sort,
for example).  The two specifications correspond to distinct three-valued
models of \texttt{merge/3} which are natural and useful.  In the first
an atom is inadmissible if \texttt{As} or \texttt{Bs} are not sorted
lists of numbers.  In the second an atom is inadmissible if \texttt{As}
or \texttt{Bs} are not lists of integers (for example, with heterogeneous
lists of integers and floats an implementation which uses the standard
ordering over terms may produce wrong answers).  Thus the different
interpretations are incomparable.  Each interpretation is a model of
many reasonable \texttt{merge/3} programs and some \texttt{merge/3}
programs have both these interpretations (and others) as models.

Ignoring the type of the list elements and comparison operator, our second
intended interpretation has more information about how the predicate
should behave.  Implementers of Prolog built-in and library predicates
surely know and care about how their predicates behave at least as much
as typical Prolog programmers, but documentation for the various versions
of \texttt{merge/3} almost always restrict attention to sorted lists
and I have never seen a more informative high level description used.
The description of \texttt{merge/3} (and many other predicates) can also
be made more precise by considering other possible modes.  If we say
that \texttt{merge/3} atoms in which the third argument is a sorted
list are also admissible we have another model, capturing the fact
that \texttt{merge/3} can work ``backwards'' to nondeterministically
split a sorted list into two, assuming a suitable computation rule,
or to check if a sorted list can be split into two given terms.
We often restrict our attention to certain modes and this leads to
less precise understanding of program behaviour from a declarative as
well as procedural view.  Similarly, most people are taught how to use
accumulators in Prolog without gaining a precise high level understanding
of the code.  Many programmers intend their \texttt{rev\_acc(As, [],
Bs)} call to succeed if and only if \texttt{Bs} is the reverse of list
\texttt{As} but only later, if ever, develop a more informed view of
the behaviour when the second argument is a non-empty list (this can
certainly be useful) or perhaps even an arbitrary term.


\section{A semantics for definite programs}

The results in this section are based on \cite{naish:sem3}.  However,
some definitions are changed slightly to make the subsequent treatment
of negation simpler.  We don't consider our semantics for definite
programs particularly useful for programmers in its own right.  However,
this section introduces several things used in our semantics for normal
programs and is useful for comparison purposes and any work on analysis
of definite programs.  We use a variant of definite clauses which we call
\emph{disjunctive definite clauses} (disjunctions only appear in the body,
not the head), primarily so conjunction and disjunction can be treated
more uniformly in the logic, though it also eases the transition to
semantics for normal programs, as does our use of an equality predicate
rather than head unification.  We also use the constraint view of
equality at some points, rather than substitutions.  This avoids some
technicalities and would make adding a ``not equals'' primitive simpler
(something definitely worth doing if negation is supported).

\begin{definition}
A \emph{disjunctive definite program} is a collection of predicate
definitions, each a single clause $H \leftarrow (B_1 \vee B_2 \vee
\ldots \vee B_n)$, where each $B_i$ is a conjunction of atoms.  $H$
is an atom $p(V_1 , V_2 , \ldots V_n )$ where the $V_i$ are distinct
variables and each $B_j$ is prefixed by a sequence of calls to a built-in
equality predicate/constraint $V_i = T_i$, where $T_i$ is a term, for each
variable $V_i$.  Other atoms in the $B_j$ are user-defined.  The $V_i$ are
called \emph{head variables}; others are called \emph{local variables}.
A \emph{head instance} of a clause is an instance where head variables
are replaced by terms and body variables are replaced by (new) distinct
variables.
\end{definition}

Disjunctive definite programs can be mapped to definite clause programs
trivially: each disjunct leads to a clause $p(T_1 , T_2 , \ldots T_n )
\leftarrow B_j'$, where $B_j'$ is $B_j$ without the initial calls to $=
/ 2$.  Definite clause programs can easily be mapped to disjunctive
definite programs by renaming variables, converting head unification
into calls to the equality predicate and combining all clauses for a
predicate into a single disjunctive clause.  This is similar to the
\emph{completion} of a program \cite{Cla78}.  We use these equivalences
implicitly when relating properties of disjunctive programs and their
Horn clause counterparts.

An intended interpretation associates each ground atom with one of three
truth values: \emph{true}, \emph{false} or \emph{inadmissible}\footnote{We
only consider \emph{consistent} interpretations; an atom cannot have
two truth values.}.  We sometimes abbreviate these to $\textbf{T}$,
$\textbf{F}$ and $\textbf{I}$ (bold font).  We also use these letters
in italics to refer to sets of user-defined atoms assigned that truth
value.  When describing interpretations we use the notation $\langle
I,T\rangle $ for the interpretation which maps the (ground) atoms in
$I$ to inadmissible, those in $T$ to true and other user-defined atoms
to false\footnote{Note this differs from the more common practice of
explicitly stating the \textbf{T} and \textbf{F} atoms.}.  Equality has
a fixed interpretation.  $E$ is the set of all ground equality atoms
of the form $X = X$.  These equality atoms are considered true; others
are considered false.  This essentially restricts interpretations to be
Herbrand interpretations.

Programmers do need to consider the fact that their programs manipulate
Herbrand terms.  It can also be useful to think at a higher level,
where terms may represent values in some other domain and two terms
may represent the same value.  However, Prolog does not respect this
form of equality.  Only syntactic equality is supported and programmers
must write explicit equality predicates or convert terms into canonical
form and be very careful with input-output modes.  Although results for
arbitrary models hold for definite programs, they typically do not hold
when negation is introduced and arbitrary models are not very useful
for programmers.

Inadmissible atoms may succeed or fail according to the procedural
semantics but this distinction is not made in the declarative semantics.
Our semantics is thus less precise than the traditional semantics,
to reflect the lack of precision of programmers.  If we consider
calls to \verb@merge/3@ where the first two arguments are sorted lists
then programmers typically know which ones should succeed.  For other
(inadmissible) calls to \verb@merge/3@, programmers typically don't
know and don't care precisely which calls succeed.  Our semantics aims
at providing tools to reason about program correctness without the
need for additional precision and without unnecessary restrictions on
programming style.  We define operational, model theoretic and fixpoint
semantics then discuss the relationships between them.

\subsection{Operational semantics}

The operational semantics are essentially the same as SLD resolution
or pure Prolog with coroutining.  Instead of a nondeterministic clause
selection followed by (possibly failed) unification and construction of
the resolvent we have nondeterministic disjunct selection followed by
(possibly failed) unification/constraint inclusion using the (multiple)
$= / 2$ calls and construction of the resolvent.

\subsection{Model-theoretic semantics}

We first define interpretations and models then discuss the motivation
for these definitions and the properties of models.

\begin{definition}
An \emph{interpretation} is a mapping from ground atoms to the truth values
\emph{true}, \emph{false} and \emph{inadmissible}.  The interpretation
of $= / 2$ is restricted to be (two-valued) syntactic equality.
\end{definition}

Conjunction and disjunction are defined as below --- the same as in
Kleene's strong three-valued logic \cite{Kleene}:


\begin{center}
\begin{minipage}{5cm}
\begin{tabular}{|c||c|c|c|}
\cline{1-4}
$\wedge$ & \textbf{T} & \textbf{F} & \textbf{I}\\
\cline{1-4}
\cline{1-4}
\textbf{T}          & \textbf{T} & \textbf{F} & \textbf{I}\\
\cline{1-4}
\textbf{F}          & \textbf{F} & \textbf{F} & \textbf{F}\\
\cline{1-4}
\textbf{I}          & \textbf{I} & \textbf{F} & \textbf{I}\\
\cline{1-4}
\end{tabular}
\end{minipage}
\begin{minipage}{5cm}
\begin{tabular}{|c||c|c|c|}
\cline{1-4}
$\vee$ & \textbf{T} & \textbf{F} & \textbf{I}\\
\cline{1-4}
\cline{1-4}
\textbf{T}        & \textbf{T} & \textbf{T} & \textbf{T}\\
\cline{1-4}
\textbf{F}        & \textbf{T} & \textbf{F} & \textbf{I}\\
\cline{1-4}
\textbf{I}        & \textbf{T} & \textbf{I} & \textbf{I}\\
\cline{1-4}
\end{tabular}
\end{minipage}


\end{center}

\begin{definition}[model of $P$]
An interpretation is a \emph{model} of a disjunctive definite program $P$
if no instance of a clause in $P$ is mapped (by the interpretation and
definitions of conjunction and disjunction) to $\textbf{F} \leftarrow
\textbf{T}$ or $\textbf{F} \leftarrow \textbf{I}$.
\end{definition}

That is, if the head is \textbf{F}, the body is \textbf{F}.  Note that the
traditional two-valued model theory is a special case of these definitions
where the set of inadmissible atoms is empty.  This definition of a
model can be seen as giving a two-valued interpretation to an implication
connective.  It differs from Kleene's (weak and strong) logics, in which
implications can have all three values.  In \cite{przymusinski89every},
and \cite{apt94logic}, an alternative definition of a model is given ---
we refer to these as \emph{strong models}:

\begin{definition}[strong model of $P$]
An interpretation is a \emph{strong model} of a disjunctive definite
program $P$ if it is a model of $P$ and no instance of a clause in $P$
is mapped to $\textbf{I} \leftarrow \textbf{T}$.
\end{definition}

Strong models significantly restrict intended interpretations and/or
programming style.  For example, they do not allow interpretations where
instances of facts in the program are inadmissible.  Our interpretation
of \texttt{merge/3} is not a strong model due to clause instances such
as \texttt{merge([],a,a)}.  To make our intended interpretation a strong
model of the first clause a test such as \verb@sorted_list(Bs)@ would have
to be added.  We have attempted to make our definition of a model as weak
as possible: it just avoids the two classes of bugs discussed earlier.
The definitions of conjunction and disjunction follow the same principle.

For conjunction, the key question is whether $\textbf{I} \wedge
\textbf{F}$ should equal $\textbf{I}$ or $\textbf{F}$.  A choice
of $\textbf{I}$ would be similar to strict type schemes in which
a conjunction containing an ill-typed atom is ill-typed, whether it
succeeds or fails.  Programs are restricted so that \emph{all} derivations
are well typed, not just successful ones (distinguishing between these
two cases cannot be done statically).  The choice of $\textbf{F}$ is
similar to less restrictive type schemes which only restrict successful
derivations; this is discussed further in \cite{naish:90}.  It allows
runtime checking of types (or other assertions) to be supported.
The body of a clause instance can have checks which fail (have truth
value $\textbf{F}$) as well as inadmissible calls, that is, the clause
is of the form $\textbf{F} \leftarrow \textbf{F} \wedge \textbf{I}$.
This would not be acceptable with the stricter definition of conjunction.

For disjunction, the key question is whether $\textbf{I} \vee \textbf{T}$
should equal $\textbf{I}$ or $\textbf{T}$.  In strict type schemes a
disjunction containing an ill-typed atom is ill-typed even if a well-typed
disjunct succeeds.  Even in the less strict scheme of \cite{naish:90}
disjunctions must be implemented as separate clauses and each clause
must be ``type correct''.  This is the reason we introduced disjunctive
clauses: by choosing $\textbf{T}$ we can have a less restrictive
condition.


The structure of the set of all models is more complex than in the
two-valued case.  There are several model intersection properties of
interest.  The first relates models with the same set of inadmissible
atoms:

\begin{proposition}
If $M_1 = \langle I,T_1\rangle $ and $M_2 = \langle I,T_2\rangle $
are models then $\langle I,T_1 \cap T_2\rangle $ is a model.
\end{proposition}
\begin{proof}
If there is a clause instance whose head is \textbf{F} in the intersection
it must be \textbf{F} in either $M_1$ or $M_2$.   Therefore the body of
the clause instance must be \textbf{F} in $M_1$ or $M_2$ (since they
are both models).  Any formula which is \textbf{F} in $M_1$ or $M_2$
must be \textbf{F} in the intersection (by inspection of the truth tables
and induction on the number of connectives).
%
\end{proof}


This proposition also holds for strong models.  It generalises the
two-valued model intersection property (where $I$ is the empty set,
$\phi$).  If there are two models with no true atoms then a similar
result holds for the intersection of sets of inadmissible atoms.
This is generalised in the following proposition (which does not hold
for strong models).

\begin{proposition}
If $M_1 = \langle I_1,T\rangle $ and $M_2 = \langle I_2,T\rangle $
are models then $\langle I_1 \cap I_2,T\rangle $ is a model.
\end{proposition}
\begin{proof}
Identical to the proof above.
\end{proof}

Models exist with no inadmissible atoms (for example, all atoms are
true) and with no true atoms (for example, all atoms are inadmissible),
so from the two propositions above it follows that a least model (with
respect to $T$) of the form $\langle \phi,T\rangle $ exists and a least
model (with respect to $I$) of the form $\langle I,\phi\rangle $ exists.
These two models have the same set of false atoms (just the role of true
and inadmissible are swapped).  These models are equivalent to the least
two-valued model of the program.

If the set of false atoms is the same in two models, either the true
atoms or inadmissible atoms can be intersected and the result is a model.
In fact, any partitioning of the true and inadmissible atoms is a model:

\begin{proposition}
If $M_1 = \langle I_1,T_1\rangle $ is a model,
$I_2 \cap T_2 = \phi$ and $T_1 \cup I_1 = T_2 \cup I_2$
then $M_2 = \langle I_2,T_2\rangle $ is a model.
\end{proposition}
\begin{proof}
Any formula which is inadmissible or true according to $M_2$ must be
inadmissible or true according to $M_1$ (by inspection of the truth
tables) so there is no clause instance of the form $\textbf{F} \leftarrow
\textbf{I}$ or $\textbf{F} \leftarrow \textbf{T}$ according to $M_2$.
\end{proof}

Thus we have a set of (at least two) minimal models (with respect to
$T$/$I$), all with the same set of false atoms (the false atoms of the
least two-valued model).

\begin{corollary}
$\langle I,T\rangle $ is a model iff $\langle \phi, I \cup T\rangle $
is a model.
\end{corollary}

This result is essentially the reason why treating inadmissible atoms
as true results in correct declarative diagnosis of wrong answers.
For models of the completion this result does not hold.

There is another natural partial ordering of three-valued interpretations
known as the information ordering (we treat the third truth value as
``don't care'', but in some contexts ``don't know'' is appropriate).

\begin{definition}[information ordering, $\subseteq_i$]
Given two interpretations, $M_1$ and $M_2$, $M_1 \subseteq_i M_2$ if $T_1
\subseteq T_2 \wedge F_1 \subseteq F_2$, where $T_1$ ($F_1$) and $T_2$
($F_2$) are the true (false) atoms in $M_1$ and $M_2$, respectively.
\end{definition}

The $\subseteq_i$-least model has all atoms inadmissible.  The
$\subseteq_i$-least strong model has the least two-valued model as the
true atoms and all other atoms inadmissible, the same as the $I$-least
model with $I$ and $F$ swapped.  Atoms in the least two-valued model
are true in all strong models.

\subsection{Fixpoint semantics}

After \cite{Fitting85} (using the terminology of \cite{apt94logic}),
we define $T3_P$, an analogue of $T_P$ which maps interpretations to
interpretations.  For simplicity, we define it in terms of the model
theory ($T_P$ could be defined in the same way --- the inadmissible
case would never arise).

\begin{definition}
$T3_P(M)$ is the interpretation such that an atom $A$ is 
\begin{enumerate}
\item
true, if there is a clause instance $A \leftarrow B$
where $B$ is true in $M$, 
\item
false, if for all clause instances $A \leftarrow B$, $B$ is false in $M$ and
\item
inadmissible, otherwise.
\end{enumerate}
\end{definition}

$T3_P$ generalises $T_P$: if there are no inadmissible atoms in $M$
there are none in $T3_P(M)$.  Various properties of $T3_P$ are discussed
in \cite{apt94logic}.  One important result is that a least fixpoint
with respect to the information ordering exists.


Another immediate consequence operator, $T3^{+}_P$, which retains the
truth value of inadmissible atoms, was introduced in \cite{naish:sem3}
(where it was called $C_P^{\prime}$).  Essentially, inadmissible atoms
are assumed to succeed.

\begin{definition}
$T3^{+}_P(M)$ is the interpretation such that an atom $A$ is
\begin{enumerate}
\item inadmissible, if $A$ is inadmissible in $M$,
\item true, if $A$ is admissible and there is a clause instance
$A \leftarrow B$ where $B$ is true or inadmissible in $M$,
\item false, otherwise.
\end{enumerate}
\end{definition}

Like $T3_P$, $T3^{+}_P$ generalises $T_P$.  If $I$ is empty, $T3^{+}_P$
is equivalent to $T_P$ and thus each fixpoint of $T_P$ corresponds
to a fixpoint of $T3^{+}_P$.  Fixpoints of $T3_P$ are also fixpoints
of $T3^{+}_P$.  Generally $T3^{+}_P$ has additional fixpoints as well.
It is particularly useful when the set of inadmissible atoms in the
intended interpretation is known.  Using $T3^{+}_P$, analysis can restrict
attention to the behaviour of admissible atoms.  If (in)admissibility can
be determined from the program, for example, from type/mode declarations,
automatic analysis may benefit.  In any case it simplifies manual analysis
and avoids analysis which can often be non-intuitive.

For example, suppose $M$ is our intended interpretation for
\texttt{merge/3}.  The inadmissible atoms in $T3^{+}_P(M)$ are the
same as those in $M$.  To find the true atoms in $T3^{+}_P(M)$ we
consider each clause and determine what \emph{admissible} atoms can
be derived in one step from the true and inadmissible atoms in $M$.
For the first two clauses we can ignore inadmissible atoms such as
\texttt{merge([],a,a)}.  For the last two clauses it is easy to show that
if the clause head is admissible then the recursive call is admissible,
so we can ignore inadmissible atoms such as \texttt{merge([2,3], [2,1],
[2,1,2,3])}.  This ``forward'' reasoning essentially replaces reasoning
about inadmissible atoms.  Forward reasoning is more procedural in nature
\cite{naish:verif:90} and essential for reasoning about instantiatedness
of calls (something logic programmers must consider).  It is simple to
establish that the true atoms in $T3^{+}_P(M)$ are a subset of those in
$M$ (in fact, they are equal).  From this we can conclude $M$ is a model
(see below).

\subsection{Relationships between semantics}
\label{sec_def_rel}

The relationship between three-valued models and $T3_P$/$T3^{+}_P$
is similar to the relationship between two-valued models and $T_P$:
\begin{proposition}
$M$ is a model iff the true atoms in $T3^{+}_P(M)$ are a subset
(or equal to) the true atoms in $M$ (that is, $T^{+} \subseteq T$
where $T3^{+}_P(\langle I, T\rangle) = \langle I, T^{+}\rangle$).
\end{proposition}
\begin{proof}

Let $T/T^{+}$ ($F/F^{+}$) be the true (false) atoms in $M/T3^{+}_P(M)$.
$M$ is a model of $P$
iff
for all clause instances $H \leftarrow B$,
$H \in F$ implies $B$ is \textbf{F} in $M$
iff
$H \in F \Rightarrow H \in F^{+}$
iff
$F^{+} \supseteq F$
iff
$T^{+} \subseteq T$
(since $T3^{+}_P$ preserves $I$).
\end{proof}

\begin{proposition}
$M$ is a model iff the false atoms in $T3_P(M)$ are a superset
(or equal to) the false atoms in $M$
(that is, $T' \cup I' \subseteq T \cup I$
where $T3_P(\langle I, T\rangle) = \langle I', T'\rangle$).
\end{proposition}
\begin{proof}
As above (change $T3^{+}_P$ to $T3_P$ and skip the last step).
\end{proof}

The relationship between least fixpoints, minimal models and success
set is more complex than the two-valued case since there are multiple
partial orders of interest and multiple minimal models by some measures.
However, nearly all minimal models we have considered bear strong
relationships with the success set $SS$ (the least fixpoint of $T_P$
and least two-valued model).  The union of true and inadmissible atoms in
models which minimise $I$ or $T$ (or $I \cup T$) is $SS$.  These models
are least fixpoints of $T3^{+}_P$ where $I$ is fixed to a suitable value
(such as $\phi$ or $SS$).  For $\subseteq_i$ the minimum strong model
has $SS$ as the true atoms and this is the least fixpoint of $T3_P$.


The main theorem we have concerns soundness for admissible atoms.
A successful atom may not be true, since inadmissible atoms can succeed.
However, an \emph{admissible} atom which succeeds must be true if the
intended interpretation is a model.

\begin{theorem}[Soundness for admissible atoms]
If $\langle I,T\rangle $ is a model, $A$ is admissible and
$A \in SS$ then $A \in T$.
\end{theorem}
\begin{proof}
$I \cup T \supseteq SS$ by the propositions above.  The result follows.
\end{proof}

A form of completeness is inherited from the two-valued case:
\begin{theorem}
\label{thm_SLDD_comp}
If $A$ is true or inadmissible in every model then $A \in SS$.
\end{theorem}
\begin{proof}
If $A$ is not false in any three-valued model it is not false in any
two-valued model (since two-valued models are three-valued models),
so it is true in every two-valued model and hence in $SS$.
\end{proof}

This result is not really useful in practice for several reasons, which
we discuss in Section\ \ref{sec_norm_rel}.

\section{A semantics for normal programs}

When negation is introduced, many of the convenient properties of the
semantics of logic programs break down.  Typically model intersection
properties no longer hold, the immediate consequence operator is no
longer monotonic and in the procedural semantics negation is associated
with \emph{finite} failure which is dependent on the computation rule.

The first and simplest approach to negation was to use the negation as
failure rule for the operational semantics and Clark's completion for
the declarative semantics \cite{Cla78}.  Here we adapt this approach to
our three-valued scheme.

\begin{definition}
\emph{Disjunctive normal programs} are the same as disjunctive definite
programs but they may have negated user-defined literals in the disjuncts.

The \emph{completion} $comp(P)$ of a disjunctive normal program $P$
is the set of completions of clauses in $P$.

The completion of a clause $H \leftarrow B$ is $H \leftarrow \exists V_1
\exists V_2 \ldots B$, where the $V_i$ are the local variables.
\end{definition}

The ``$\leftarrow$'' in completions is interpreted as ``if and only
if'', modulo inadmissible atoms, in the model theory (see below).
Clause instances of the form $\textbf{T} \leftarrow \textbf{F}$ are
a source of incompleteness (a class of bugs called uncovered atoms
\cite{Sha83}) which can translate to unsoundness when negation as failure
is used.  Clause instances of the form $\textbf{T} \leftarrow \textbf{I}$
can cause similar problems.  Models of the completion therefore do
not include such clauses.  Clause instances of the form $\textbf{I}
\leftarrow \textbf{F}$ and $\textbf{I} \leftarrow \textbf{T}$ do not
to cause problems for Prolog.  Even alternative execution mechanisms
would not cause problems as long as the top level of the computation is
assumed to be admissible or any inadmissible instances can be ignored.
Due to the asymmetry with \textbf{I} we retain $\leftarrow$ instead of
using $\leftrightarrow$.

One criticism of Clark's approach is that the completion may have no
(two-valued) models. Adding a definition such as $p \leftarrow \neg p$
to a program results in everything being a logical consequence of the
program but does not change the success set.  Though we believe this
criticism has been over-stated, it does not apply in the three-valued
case.  A three-valued model (and even strong model) always exists and an
atom such as $p$ can be inadmissible, independent of the interpretation
of the rest of the program.

\subsection{Operational semantics}

The operational semantics we use is essentially SLDNF resolution,
where negative literals must be ground in order to be selected.
We define trees to formalise the operational semantics.  There are a
few differences between our definitions and the standard SLDNF tree
definitions.  First, for technical convenience evaluation of negative
literals is done within the same tree structure (like \cite{martelli},
rather than having separate trees of different ranks) and equality
atoms/constraints are used rather than substitutions.  Second, treatment
of floundering within negation is improved (it is often done poorly).
Third, we distinguish between searching for all solutions and just some
solution(s).  Our main aim is to establish results about observables from
Prolog computations, namely, zero or more (possibly floundered) computed
answers and possibly an indication there are no (more) answers (we ignore
computations which are aborted --- we have no results for such cases).
These are always the result of a finite search and we define finite
trees which correspond to such computations.  Even if all SLDNF trees
are infinite the search may be finite because only some solution(s)
may be needed (at the top level or inside a negation).

\begin{definition}[SLDDNF tree]
An SLDDNF tree is a (possibly infinite) tree where nodes are connected
by positive or negative edges.  The positive nodes of a (sub)tree are
those connected to the root with a sequence of positive edges.

Each node contains a conjunction of literals including equality atoms.
Nodes containing an unsatisfiable set of equality atoms are said to be
\emph{failed} and have no children.  Nodes containing a satisfiable set
of equality atoms and no other literals are said to be \emph{successful}
and have no children.  A literal is said to be \emph{grounded} if the
substitution obtained by unifying the arguments of each equality atom
would make the literal ground.  Nodes containing only a satisfiable
set of equality atoms and non-grounded negative literals are said
to be \emph{floundered} and have no children.  Other nodes have a
\emph{selected literal}, which is not an equality atom and must be
grounded if it is negative.

If the goal in node $N$ is $L_1 \wedge \ldots L_n$ and the selected
literal is $L_m$, then
\begin{itemize}

\item If $L_m$ is an atom $A$ and $A \leftarrow B_1 \vee B_2 \ldots
B_k$ is a head instance of its definition, then $N$ has $k$ children,
connected with positive edges, with goals $L_1 \wedge \ldots L_{m-1}
\wedge B_i \wedge L_{m+1} \ldots L_n$, for $1 \leq i \leq k$.

\item If $L_m$ is a negative literal $\neg A'$, there is one child
connected with a negative edge, containing goal $A'$ and the equality
atoms from $N$.  If the subtree for $A'$ has a positive successful node
then $N$ has a child, connected with a positive edge, which is failed.
If the subtree for $A'$ is finite and all positive leaves are failed,
then $N$ has a child, connected with a positive edge, which contains $L_1
\wedge \ldots L_{m-1} \wedge L_{m+1} \ldots L_n$.  If the subtree for $A'$
is finite, there are no positive successful nodes but there is a positive
floundered node, then $N$ has a child, connected with a positive edge,
which is identical to $N$.  Otherwise (the negative subtree is infinite
with no successful nodes), $N$ has no positive children and is considered
a positive leaf.
\end{itemize}
\end{definition}


\begin{definition}[Observations tree]
An (SLDDNF) observations (sub)tree $O$ is a \emph{finite} subset of the
nodes/branches of an SLDDNF tree $S$ such that
\begin{enumerate}
\item the leaves of $O$ are leaves of $S$,
\item if $O$ has no positive successful leaves it has all positive
nodes of $S$, and
\item for each selected negative literal in $O$ there is an observations
subtree of the corresponding subtree in $S$.
\end{enumerate}
\end{definition}

\begin{definition}[All-observations tree]
An (SLDDNF) all-observations tree is an observations subtree of
SLDDNF tree $S$ which includes all positive nodes of $S$.
\end{definition}

A Prolog implementation can be seen as searching an SLDDNF tree (typically
depth-first and left to right) for one or more successful positive nodes.
When such nodes are found at the top level the equations in the node
(equivalent to variable bindings) may be displayed in a suitable
fashion and the search may stop.  When such nodes are found within a
negation the search typically stops and backtracking is initiated at
the higher level (where the negation was called).  Observation trees
can model such behaviour.   All-observations trees model computations
which find all solutions and terminate.  Finitely failed observations
trees are all-observations trees (note that corresponding SLDDNF trees
may have infinite branches inside negations).  We do not explicitly
model computations which flounder without succeeding or searching the
entire tree.  They are of limited interest, especially inside negation,
though it would be easy to modify our definitions to support them.
Similarly, we do not model non-depth-first computations where some
branches are only partially searched.

Many implementations neglect to check that negative literals are ground
(leading to unsoundness), and even those which do typically have unsound
treatment of floundering within negation (this is sometimes treated
poorly in the theoretical literature also).  Our (novel) solution here
is that selecting a negative literal which flounders does nothing to
the current goal.  If a different literal is selected subsequently,
which would occur with a fair computation rule, failure may result; if
the same literal is always selected the tree will be infinite.  If $N$
was considered a positive leaf in this case instead, and an exception
mechanism was invoked or some kind of abnormal termination was flagged
it could be more practical, but harder to formalise.

\begin{figure}
\figrule
\begin{verbatim}
p :- not q.
q :- not r, not s.
r :- not t(_).
s.
\end{verbatim}
\caption{Literal selection and floundering}
\figrule
\label{fig_litsel}
\end{figure}

An advantage of our approach (at least in theory) is that it potentially
avoids a source of incompleteness.  Suppose we have a goal with two ground
negated atoms, one of which flounders and the other succeeds.  There is
no \emph{a priori} way of determining which literal should be selected.
For example, in Figure\ \thefigure, the goal $\leftarrow p$ has a
successful SLDNF derivation but when the resolvent $\leftarrow \neg
r, \neg s$ is encountered the right literal must be selected to avoid
floundering using the normal semantics.  With our semantics we may first
select $\neg r$ but that just leaves the current goal unchanged, allowing
us to then select $\neg s$ (a fair computation rule would select $\neg
s$ eventually; an unfair rule may result in a loop).

\subsection{Model-theoretic semantics}
\label{sec_norm_mod}

We define models for the completion of a program.  First, negation is
defined by $\neg \textbf{T} = \textbf{F}, \neg \textbf{F} = \textbf{T},
\neg \textbf{I} = \textbf{I}$.  The existential quantification of a closed
formula is $\textbf{T}$ if any instance is $\textbf{T}$, $\textbf{F}$
if all instances are $\textbf{F}$, and $\textbf{I}$ otherwise.

\begin{definition}[model of $comp(P)$]
An interpretation is a \emph{model} of $comp(P)$ if it is a model of
every clause in $comp(P)$.

An interpretation is a \emph{model} of $H \leftarrow B$ if for all
head instances $H \theta \leftarrow B \theta$, if $H \theta$ is
\textbf{T} then $\exists(B \theta)$ is \textbf{T} and if $H \theta$ is
\textbf{F} then $\exists(B \theta)$ is \textbf{F} in the interpretation.
\end{definition}

That is, we avoid clause instances of the form
$\textbf{F} \leftarrow \textbf{T}$,
$\textbf{F} \leftarrow \textbf{I}$,
$\textbf{T} \leftarrow \textbf{F}$ and
$\textbf{T} \leftarrow \textbf{I}$.
The first two cases can cause unsoundness for definite clauses, as
discussed earlier.  The last two cases can cause unsoundness with
negation as failure. Note that we only consider instantiation of head
variables and use existential quantification in the bodies.  We allow
the case where the instance of $H$ is $\textbf{T}$ and some but not
all corresponding instances of $B$ are $\textbf{T}$ (corresponding
to a true atom with a legitimate proof and one or more suspect proofs
which use inadmissible or false atoms).  Adapting the definite clause
model definition in a simpler way results in a stronger definition of a
model, unnecessarily rejecting some programs for a given interpretation.

To summarise, the declarative semantics we propose for logic programs is
Clark's completion with Kleene's strong three-valued logic used for the
right sides of the arrow and the following truth table used for the arrow.


\begin{tabular}{|c||c|c|c|}
\cline{1-4}
$\leftarrow$ & \textbf{T}  & \textbf{F} & \textbf{I} \\
\cline{1-4}
\cline{1-4}
\textbf{T}   & \textbf{T}  & \textbf{F} & \textbf{F} \\
\cline{1-4}
\textbf{F}   & \textbf{F}  & \textbf{T} & \textbf{F} \\
\cline{1-4}
\textbf{I}   & \textbf{T}  & \textbf{T} & \textbf{T} \\
\cline{1-4}
\end{tabular}\\

The model intersection properties stated earlier do not generally hold
for disjunctive normal programs.  We define strong models of completions
in the typical way (see \cite{apt94logic}), additionally avoiding clauses
of the form $\textbf{I} \leftarrow \textbf{T}$ and $\textbf{I} \leftarrow
\textbf{F}$:

\begin{definition}[strong model of $comp(P)$]
An interpretation $M$ is a \emph{strong model} of $comp(P)$ if every head
clause instance has the same truth value for the head and body in $M$.
\end{definition}

Our intended interpretation of \texttt{merge/3} is not a strong model
even if extra tests are added.  Clauses such as \texttt{merge([], Bs,
Bs) :- sorted\_list(Bs)} have instances where the head is \textbf{I}
and the body is \textbf{F}.  Strong models must precisely specify the
behaviour of all predicates.

\subsection{Fixpoint semantics}

The $T3_P$ definition given earlier can be applied when negative
literals are present.  It is equivalent to defining $T3_P(M)$ to be the
interpretation such that the truth value for each atom $A$ is the truth
value of $B$ in $M$, where $A \leftarrow B$ is a head clause instance in
$comp(P)$.  In the presence of negation $T3_P$ is generally not monotonic
with respect to $T$ but is monotonic with respect to the information
measure and, unlike $T_P$, at least one fixpoint exists and the
$\subseteq_i$-least fixpoint can be built using $T3_P$ \cite{Fitting85}.

For definite programs we introduced $T3^{+}_P$, which essentially
overestimates the set of successful atoms.  This is so even when negation
is present --- a negated inadmissible atom is inadmissible and hence
assumed to succeed (the negation makes no difference).  When negation
is present it is helpful to also have an operator which underestimates
this set, by assuming inadmissible clause body instances fail:

\begin{definition}
$T3^{-}_P(M)$ is the interpretation such that an atom $A$ is
\begin{enumerate}
\item inadmissible, if $A$ is inadmissible in $M$,
\item true, if there is a clause instance $A \leftarrow B$ where $B$ is
true in $M$,
\item false, otherwise.
\end{enumerate}
\end{definition}

As with $T3^{+}_P$, $T3^{-}_P$ generalises $T_P$ and its fixpoints
include all those of $T_P$ and $T3_P$.

\subsection{Relationships between semantics}
\label{sec_norm_rel}

For normal programs, fixpoints of $T_P$ correspond to (two-valued)
Herbrand models of $comp(P)$.  For each such fixpoint there are
corresponding fixpoints of $T3_{P}$, $T3^{-}_{P}$ and $T3^{+}_{P}$
where $I$ is empty, and a corresponding three-valued model of $comp(P)$.
The following propositions generalise this result to the case where $I$
may be non-empty.

\begin{proposition}
$M$ is a model of $comp(P)$ iff $T3^{+}_P(M) = M$ and $T3^{-}_P(M) = M$.
\end{proposition}
\begin{proof}
Let $T/T^{+}/T^{-}$ ($F/F^{+}/F^{-}$) be the true (false) atoms in
$M/T3^{+}_P(M)/T3^{-}_P(M)$.
$M$ is a model of $comp(P)$
iff
for all head clause instances $H \leftarrow B$, $H \in T$ implies
$B$ is \textbf{T} in $M$ and $H \in F$ implies $B$ is \textbf{F} in $M$
iff
$H \in T \Rightarrow H \in T^{-}$ and $H \in F \Rightarrow H \in F^{+}$
iff
$T^{-} \supseteq T \wedge F^{+} \supseteq F$
iff
$T = T^{-} \wedge F = F^{+}$
(since $T \supseteq T^{-}$ and $F \supseteq F^{+}$)
iff
$M$ is a fixpoint of $T3^{+}_P$ and $T3^{-}_P$
(since these operators preserve $I$).
\end{proof}

\begin{proposition}
$M$ is a model of $comp(P)$ iff
$ M \subseteq_i T3_P(M)$.
\end{proposition}
\begin{proof}
Let $T$ ($F$) and $T'$ ($F'$) be the true (false) atoms in $M$ and
$T3_P(M)$, respectively.  For a head clause instance $H \leftarrow B$,
the truth value of $H$ in $T3_P(M)$ is the truth value of $B$ in $M$.
$T3_P(M) \subseteq_i M$ iff
$F' \supseteq F \wedge T' \supseteq T$ iff
$H \in F$ implies $H \in F'$ and $H \in T$ implies $H \in T'$ iff
$H$ is \textbf{F} in $M$ implies $B$ is \textbf{F} in $M$ and
$H$ is \textbf{T} in $M$ implies $B$ is \textbf{T} in $M$ iff
$M$ is a model.
\end{proof}

\cite{apt94logic} gives a detailed account of the relationships between
strong models, $T3_{P}$ and the operational semantics.  Strong models
coincide with fixpoints of $T3_{P}$ and the $\subseteq_i$-least strong
model and fixpoint captures the operational semantics \cite{Fitting85}.
The true atoms in this model are those in $SS$ and the false atoms
are those in $FF$, the atoms with finitely failed SLDNF trees.
\cite{Kun87} established conditions under which these sets of atoms are
also the two-valued logical consequences of the program.  This fixpoint
characterisation of the operational semantics is potentially very useful
for program analysis and alternative operational semantics (bottom-up
execution).  However, the model theory is not particularly helpful for
programmers to reason about correctness or debug their programs.

In all strong models, successful atoms are true, finitely failing atoms
are false and thus only looping atoms can be inadmissible.  It is not
possible to under-specify the behaviour of predicates as we can using
our definition of a model.  Fifteen of our versions of even and odd
terminate for all ground queries so the intended interpretation is not a
strong model.  However, over-specification is possible in the sense that
looping atoms can be true or false in strong models.  For the version of
even and odd where everything loops the intended interpretation \emph{is}
a strong model.


The following lemma relates our model theoretic semantics and the
operational semantics, essentially establishing soundness and an important
form of completeness.  With our definition of models of clauses, truth
and falsity of clause heads is propagated to clause bodies.  The lemma
shows they propagate from the root to positive leaves of SLDDNF trees
(in the other direction, inadmissibility can be introduced).  This is
the contrapositive of the normal statement of results such as soundness.

\begin{lemma}
\label{lem_SLDDNF}

Suppose $M$ is a model of $comp(P)$, $G$ a goal, $S$ an SLDDNF tree
for $\leftarrow G \cup P$, $O$ an observations subtree of $S$ and
$\theta$ a substitution of terms for the variables in $G$.
If $G \theta$ is ground and \textbf{T}
and $O$ is an all-observations subtree of $S$
with positive leaves $L_1, L_2, \ldots$ then
$\exists (L_1 \theta \vee L_2 \theta \vee \ldots)$ is \textbf{T}.
If $\exists(G \theta)$ is \textbf{F} and $O$ has a positive leaf $L$ then
$\exists (L \theta)$ is \textbf{F}.
\end{lemma}
\begin{proof}

We use induction on the height of $O$.
For height zero ($G$ is a positive leaf) it is trivial.

Assume the lemma holds for (sub)trees of height $\leq n$; we show
it holds for height $n+1$.  Without loss of generality we assume $G$
(which has satisfiable equality constraints) may be reordered to the
conjunction $A \wedge R$, where $A$ is the selected literal.


If $A$ is an atom, defined by $A \leftarrow D_1 \vee D_2 \ldots$,
the children are $D_i \wedge R$ if $O$ is an all-observations tree
(otherwise they are a non-empty subset of these).

Suppose $G \theta$ is \textbf{T}, $O$ is an all-observations tree and the positive
leaves of all the $D_i \wedge R$ subtrees are $L_{s_i}, L_{s_i +1} \ldots
L_{f_i}$.  $R \theta$  and $A \theta$ are \textbf{T} and some $\exists
(D_i \theta)$ is \textbf{T} (since $M$ is a model).  So an instance of
$D_i \theta \wedge R \theta$ is \textbf{T} so $\exists (L_{s_i} \theta
\vee \ldots L_{f_i} \theta)$ is \textbf{T} (by the induction hypothesis)
so $\exists (L_1 \theta \vee L_2 \theta \vee \ldots)$ is \textbf{T}.
Suppose $\exists (G \theta)$ is \textbf{F}.  For any ground instance
$G \theta \phi$, either $R \theta \phi$ is \textbf{F} or both $A \theta
\phi$ is \textbf{F} and all $\exists (D_i \theta \phi)$ are \textbf{F}
(since $M$ is a model).  So for every $i$, $\exists (D_i \theta \wedge
R \theta)$ is \textbf{F} and by the induction hypothesis all instances
of positive leaves are \textbf{F}.  Thus, if an atom is selected the
lemma holds for a tree of height $n+1$.

If $A$ is a negated atom $\neg A'$, the equality constraints of $G$ are
$C$ and $\gamma$ is the grounding substitution for $A'$ derived from $C$
(note all instances of $C \gamma$ are \textbf{T}), the negative subtree
may be successful, unsuccessful but finitely floundered or finitely
failed.  In the floundering case there is a single positive child with
goal $G$ and by the induction hypothesis the lemma holds.


If the negative subtree has a positive successful leaf $L$ then there is
a single positive child of $G$ which is failed.  It is sufficient to show
that no instance of $G$ is \textbf{T}.  $C$ is a subset of the equality
constraints in $L$ and $\exists (L \gamma)$ is \textbf{T}, so not all
instances of $A' \gamma \wedge C \gamma$ are \textbf{F} (by the induction
hypothesis).  All instances of $C \gamma$ are \textbf{T} so $A' \gamma$
(which is ground) is not \textbf{F}.  Thus for all instances of $G$,
either $\neg A'$ is not \textbf{T} or $C$ is \textbf{F} so no instance
of $G$ is \textbf{T}.



If the negative subtree is finitely failed then it is an all-observations tree
and there is a single positive
child with goal $R$.  By the induction hypothesis it is sufficient to
show that if $G \theta$ is \textbf{T} then $R \theta$ is \textbf{T}
and if $\exists (G \theta)$ is \textbf{F} then $\exists (R \theta)$
is \textbf{F}.  The first case is straightforward; we prove the second
by contradiction.  If $\exists (R \theta)$ is not \textbf{F} but $\exists
(\neg A' \theta \wedge R \theta)$ is \textbf{F} then there must be a
substitution $\phi$ such that the set of constraints $C \theta \phi$ is
\textbf{T} and $\neg A' \theta \phi$ is \textbf{F}.  Thus $(A' \wedge
C) \theta \phi$ is \textbf{T}.  But $A' \wedge C$ is finitely failed,
so by the induction hypothesis has no instance which is \textbf{T}.
\end{proof}

\begin{theorem}[soundness modulo inadmissibility]
If $M$ is a model of $comp(P)$, $G$ is a goal and an SLDDNF observations
tree for $G$ and $P$ of has a positive successful leaf $L$ then any ground
instance $G \theta$ consistent with $L$ (that is, $\exists (L \theta)$
is \textbf{T}) is true or inadmissible in $M$.
\end{theorem}
\begin{proof}
The constraints in all ancestors of $L$ are consistent with $\theta$ since
they are subsets of $L$.  Thus an observations tree for $G \theta$ using the
same computation rule has a positive leaf $L \theta$, which is successful.
Thus, by Lemma \ref{lem_SLDDNF}, $G \theta$ is not false in any model.
\end{proof}

If we consider computed answers being returned rather than equality
constraints, this theorem tells us that any instance of a computed answer
is true or inadmissible in every model.

\begin{theorem}[soundness of finite failure]
If $M$ is a model of $comp(P)$, $G$ is a goal and an SLDDNF observations
tree for $G$ and $P$ is finitely failed then no instance of $G$ is true
in $M$.
\end{theorem}
\begin{proof}
If $G \theta$ is \textbf{T} in $M$ then by Lemma \ref{lem_SLDDNF}
the existential closure of the leaves is \textbf{T}, but the tree is
finitely failed so this cannot be the case.
\end{proof}

The following completeness result is of significant practical use to
programmers (as are similar results for the two-valued case we have
stated in the past).  By completeness we mean lack of missing answers
in an ``all solutions'' computation which terminates normally (like
\cite{drabentTPLP}) rather than the existence of a tree (which may
or may not be found in practice).  It could also be seen as a form of
``all solutions'' soundness.

\begin{theorem}[strong completeness for all-observations trees]
If $M$ is a model of $comp(P)$, $G$ is a goal and a SLDDNF
all-observations tree for $G$ and $P$ has positive leaves $L_1, L_2,
\ldots$ then any instance $G \theta$ which is true in $M$ is consistent
with some $L_i$ (that is, $\exists (L_i \theta)$ is \textbf{T}).
\end{theorem}
\begin{proof}
Follows from Lemma \ref{lem_SLDDNF}.
\end{proof}


Many other completeness results (for example, our Theorem
\ref{thm_SLDD_comp}) are far less useful for programmers for three
reasons.  First, even if some proof strategy is complete in theory,
generally the completeness is dependent on forms of fairness (both
computation rule and search strategy) which are not adhered to in
implementations.  Second, even with a fair implementation we cannot
rely on a proof being found in practice due to resource limits and
other runtime errors (fair search makes this problem worse because
algorithms with acceptable complexity become extremely difficult
to express).  Third, completeness results often concern atoms which
are true in \emph{all} models (with a suitable definition of a model).
Though a programmer may know an atom is true in their \emph{intended}
interpretation (and be confident that the interpretation is a model),
knowing it is true in all models is unlikely.  In fact, the simplest way
for a programmer to be confident something is true in all models is by
reasoning about termination of the operational semantics.  Similarly,
we are not convinced that many programmers find it simple or natural
to reason about the least model or all minimal models or well-founded
models or perfect models or stable models.




In fifteen of the sixteen versions of \texttt{even/1} and \texttt{odd/1}
it is very easy to reason that ground queries terminate.  It is also
easy to show our intended interpretation is a model, by checking one
clause at a time.  From our completeness result we can conclude that
goals which are true in our intended interpretation will be successful,
assuming there are enough resources at runtime.  No other completeness
result gives us this information.  They rely on fairness and models
which are not our intended interpretation.  Some results do not apply
to all versions (for example, not all versions are stratified).

The relationship between the operational semantics and various forms of
model-theoretic semantics can be summarised by the following table:

\begin{tabular}{|l|c|c|c|}
\cline{1-4}
Operational semantics & may succeed & must loop & may finitely fail \\
\cline{1-4}
$\subseteq_i$-least strong model & \textbf{T} & \textbf{I} & \textbf{F} \\
\cline{1-4}
any strong model & \textbf{T} &
		\textbf{T}/\textbf{I}/\textbf{F} & \textbf{F} \\
\cline{1-4}
any model & \textbf{T}/\textbf{I} &
		\textbf{T}/\textbf{I}/\textbf{F} & \textbf{I}/\textbf{F} \\
\cline{1-4}
\cline{1-4}
\end{tabular}\\

We have a simplified view of the operational semantics for comparison
purposes.  Success and finite failure are generally conditional on
fairness; looping may occur with an unfair search or computation rule.
Looping may also occur for successful queries if all solutions are sought.
Some queries cannot succeed or finitely fail, even with fairness, and are
classified as ``must loop''.  We have also ignored floundering, which
can occur as well as or instead of the other behaviours.  For example,
atoms which are \textbf{T} in the $\subseteq_i$-least strong model may
actually flounder instead of succeeding.

Compared to the first two semantics, the last two require much simpler
reasoning for the programmer to establish correctness.  The last
semantics (which is what we propose) is less precise than the strong model
semantics.  However, the difference in precision is only for inadmissible
atoms.  If we know an atom is \textbf{T} or \textbf{F} the strong model
semantics gives us no additional information about how the atom behaves.
Since we don't care about the behaviour of inadmissible atoms the lack
of precision is of no concern, but the greater flexibility (existence
of more models) is very useful for allowing more natural interpretations
without restricting programming style.

\section{Program verification}
\label{sec_verif}

\begin{figure}
\figrule
\begin{verbatim}
subset(L, M) :- not notsubset(L, M).

notsubset(L, M) :- member(X, L), not member(X, M).

member(X, [X|L]).
member(X, [Y|L]) :- member(X, L).
\end{verbatim}
\caption{Definition of \texttt{subset/2}}
\figrule
\label{fig_subset}
\end{figure}

One of the main motivations for our work is program verification and from
this viewpoint it is very similar to our earlier work \cite{naish:90}
and \cite{drabentTPLP} (which contain more references related to
this area).  All assume that for correct programs, some atoms should
succeed, some should fail and for some we don't care.  The soundness
and completeness results of \cite{drabentTPLP} are very similar
to ours, including the treatment of non-termination; the verification
methods establish the same program properties.  The main thing
that distinguishes our current approach is the \emph{explicit} use of
three-valued logic.  We discuss the two examples of verifying definitions
of the subset relationship, where sets are represented as lists, used in
\cite{drabentTPLP} (and elsewhere).  For the first example, given
in Figure\ \thefigure, we describe how our verification method could
proceed and compare it with two other methods.  For the second example we
just describe our method.  The intended interpretations we use for
verification are chosen to be consistent with \cite{drabentTPLP}.
We also discuss other possible intended interpretations later.  This gives
some new insights into the relationship between the two examples and
highlights the fact that programs can have more than one natural meaning.

With our approach to verification of Figure\ \thefigure, we would use the
following definition of admissibility: both arguments of \texttt{subset/2}
and \texttt{notsubset/2} and the second argument of \texttt{member/2}
are lists.  The true and false atoms should be clear from the predicate
names and description above.  By showing this interpretation is a model
of the program, our soundness and completeness results can be established
for this program.

We can show our interpretation is a model of the definition of
\texttt{notsubset/2}, for example, by the following reasoning.
If \texttt{notsubset(L,M)} is \textbf{T}, \texttt{L} and \texttt{M}
are lists and there is an element of \texttt{L} which is not an
element of \texttt{M}, so the body of the clause is \textbf{T}.
If \texttt{notsubset(L,M)} is \textbf{F}, \texttt{L} and \texttt{M}
are lists and there is no element of \texttt{L} which is not an element
of \texttt{M}, so the body of the clause is \textbf{F}.  The reasoning
for the definition of \texttt{subset/2} is trivial since the \textbf{T}
(\textbf{F}) atoms of \texttt{subset/2} are the \textbf{F} (\textbf{T})
atoms of \texttt{notsubset/2}.  As well as such direct proofs we could
apply the immediate consequence operators and use the propositions
relating them to models.

In \cite{naish:90} (our first approach to dealing with inadmissibility)
there are two stages to verification.  The first is to show that
well-typedness (admissibility) is propagated from clause heads to clause
bodies.  It is similar to showing that if the clause head is \textbf{T}
or \textbf{F} the body is \textbf{T} or \textbf{F} (the definition is
rather more complex, restricting attention to instances of negated atoms
and the clause body which are true in some model).  The second is to show
the intended interpretation, where inadmissible atoms are considered
\textbf{F}, is a (two-valued) model of a modified version of the
program with ``type checks'' added to clause bodies (so that the body is
\textbf{F} if the head is inadmissible).  To verify \texttt{notsubset/2}
and \texttt{subset/2}, showing admissibility is propagated from heads to
bodies is straightforward.  Checking the interpretation is a model of the
modified program is similar to the three-valued approach (though there
are two extra calls in the clause bodies and instances where \texttt{L}
or \texttt{M} are non-lists must be considered as well).  Overall, the
method is more complicated than what we now propose, and models other
than the intended interpretation must be considered at one point.

In \cite{drabentTPLP}, two two-valued ``specifications'' are used, as
discussed in Section\ \ref{sec_bg_neg}.  The ``completeness
specification'' contains what we refer to as true atoms and the
``soundness specification'' contains the true and inadmissible atoms.
A diagram clearly shows the three truth values we use in intended
interpretations and an earlier version of the paper,
\cite{drabent01proving}, stated the pair of specifications ``is a
formalisation of such interpretations''.
Each specification is simpler than our
three-valued interpretations but because two specifications are always
required it is more complicated overall.  There are four sets (two
partitionings of the Herbrand base) instead of our three (a three-way
partitioning).  To support negation using this encoding of three values
there is a primed (as well as original) version of each predicate, also
increasing complexity.  There is a priming operation on specifications
and priming and double priming operations on (sets of) formulas which
convert (some) atoms in the specifications/formulas into primed versions.

To verify the \texttt{subset/2} definition a mixture of unions of
sometimes primed specifications and sometimes primed formulas is used
\cite{drabent01proving}.
This is a very contorted way of getting at the truth table for negation
in the three-valued logic, which is what our verification method uses
directly.  For \texttt{notsubset/2} there are similar contortions but at
the core of the proof there is identical reasoning to the \textbf{T} case
in our three-valued approach.  Our \textbf{F} case is done by showing
that if the body is \textbf{T} or \textbf{I}, the head is \textbf{T}
or \textbf{I}.  Though slightly more complex, this may be more natural
for many programmers.  We feel that explicit three valued logic makes
the equivalence of the two more obvious, making it easier to choose the
method that seems most natural.

\begin{figure}
\figrule
\begin{verbatim}
subs([], L).
subs([H|T], LH) :- select(H, LH, L), subs(T, L), not member(H, T).

select(H, [H|L], L).
select(H, [X|L], [X|LH]) :- select(H, L, LH).
\end{verbatim}
\caption{Alternative definition of subset, \texttt{subs/2}}
\figrule
\label{fig_subs}
\end{figure}

Figure\ \thefigure\  gives an alternative definition of the subset
relationship, which can be used to generate subsets rather than only
test them, so it better illustrates completeness of (terminating)
all-solutions computations.  Lists containing duplicates must be
avoided in some places to make (finite) generation of subsets possible.
Our intended interpretation is as follows (for \texttt{member/2} it is
the same as earlier).
\begin{itemize}
\item
\texttt{subs(L,M)} is admissible if \texttt{M} is a list; it is true if
\texttt{L} is a duplicate-free list whose elements are a subset of those 
in \texttt{M}.
\item
\texttt{select(E,L,M)} is admissible if \texttt{L} is a list; it is true if
\texttt{M} is a list the same as \texttt{L} but with one extra element
\texttt{E} at some point.    
\end{itemize}

For completeness we must show every true atom matches with a ground
clause instance with a true body.  If \verb@subs(L,LH)@ is \textbf{T},
then either \texttt{L} is \verb@[]@ (and the first clause matches) or is
of the form \verb@[H|T]@.  If \verb@subs([H|T],LH)@ is \textbf{T}, then
\texttt{H} is not a member of \texttt{T} (since it is duplicate-free),
but is a member of \texttt{LH}, which is a list, so there exist \texttt{L}
such that \texttt{select(H,LH,L)} and \texttt{subs(T,L)} are \textbf{T}.
That is, there is a matching instance of the second clause with a
\textbf{T} body.

For soundness we must show every false atom matches only with
ground clause instances with false bodies.  \verb@Subs([],L)@ cannot
be \textbf{F}.  If \verb@subs([H|T],LH)@ is \textbf{F}, then either
\verb@[H|T]@ is not a duplicate-free list or its elements are not a subset
of those in \texttt{LH}.  That is, \texttt{H} is a member of \texttt{T}
or \texttt{T} is not a duplicate-free list or \texttt{H} is not a member
of \texttt{LH} or a member of \texttt{T} (other than \texttt{H}) is not
a member of \texttt{LH}.  That is, \texttt{not member(H,T)} is \textbf{F}
or for all \texttt{L}, the conjunction \texttt{select(H,LH,L), subs(T,L)}
is \textbf{F}.

The interesting relationship between the two subset programs can be
clarified by examining other (three-valued) models of the programs.
Although the success sets and the intended models we have described
above are incomparable, we know both programs would be acceptable for
many applications and the second program is more flexible in terms of
modes (the first argument is not required to be input).  One important
difference between the programs is the restriction on duplicates in
lists introduced.  We can have a different intended interpretation of the
first program where \texttt{subset/2} atoms containing duplicates in the
first argument are \textbf{I}.  This interpretation is a model and the
\textbf{T} atoms are the same as those in our intended interpretation of
\texttt{subs/2}.  The key difference between these two interpretations
is that some atoms are \textbf{I} for \texttt{subset/2} but \textbf{F}
for \texttt{subs/2}.

The interpretation for \texttt{subset/2} has \emph{less information}
($\subseteq_i$) than that for \texttt{subs/2}.  The interpretation with less
information is a model of both programs but the interpretation with more
information is only a model of the program with more flexible modes.
Programs which use a subset predicate only as a test can be verified
using either interpretation, whereas programs which require generation
of subsets need the more precise interpretation and the \texttt{subset/2}
definition would not be acceptable.  Similarly, some programs use a subset
predicate with only duplicate-free lists in the \emph{second} argument.
This leads to two more interpretations which are less precise but
arguably more intuitive and have the same properties as the previous two.
Relationships between modes and two-valued models of definite programs,
and how they can be used to verify certain properties of programs,
are discussed in \cite{modes}.  We believe three-valued models may be
useful for extending this work.  For example, increasing flexibility
due to the existence of more models, and supporting negation.

%
%

\section{Conclusion}

In the early days of logic programming, much was made of the closeness of
logic programs and specifications.  Some people went as far as saying they
were the same; others suggested logic programs were logical consequences
of specifications.  One of the failings of this work was the lack of
recognition that specifications don't typically define what is correct
behaviour in all cases, whereas the two-valued declarative semantics
of logic programs must.  The use of classical logic instead of Kleene's
strong three-valued logic as the starting point for the declarative
semantics was, we believe, a technical mistake (though it probably
helped the early popularity of logic programming).  What was once an
important selling point of logic programming has been largely discounted
in recent times, but our work shows the relationship between logic programs
and specifications \emph{is} much closer than we have come to accept.

Unfortunately, there tends to be strong resistance to non-classical logics
(having been part of the resistance in the past I'm now a collaborator).
A typical Prolog programmer told ``if you learn about a variation of
Kleene's strong three-valued logic applied to Clark's completion you
will be able to verify your programs more easily'' is unlikely to jump
at the opportunity.  Although the statement is true and the three-valued
approach is significantly simpler than the alternatives, it does pose
an educational challenge.  It may be possible to finesse this problem by
describing techniques programmers can use without using technical jargon,
something we have attempted to do.

Our work was motivated by the desire to find a better answer to the
following question.  \emph{What is the weakest condition a programmer
should enforce which ensures correct behaviour of their programs?} In the
current context we identify correct behaviour primarily with soundness and
a form of completeness, though in general, termination, other forms of
completeness and even efficiency are important.  The traditional answer
is that the programmer should have a two-valued interpretation which is a
model of the program (or completion).  A strictly weaker condition is
that the programmer should have a three-valued interpretation which is a
(strong) model.  Our answer is similar, but strictly weaker again due
to our definition of a model.  This allows more natural interpretations
without unduly restricting programming style. It is consistent with
declarative debugging and other approaches to verification of logic
programs which are not explicitly based on model theory.

From a more technical perspective, the model theory seems quite elegant.
The fixpoint theory is more complex than the two-valued case but retains
important relationships with the model theory.  Our semantics is very
programmer-oriented, giving no immediate help to those interested in
automatic program analysis.  However, various declarations concerning
types, modes, assertions, \emph{et cetera} can be seen as documenting
(an approximation to) the set of inadmissible atoms in the intended
interpretation and can be used for program analysis and optimisation.
Such declarations are thus compatible with our semantics but we have
argued that no formalism will precisely capture the intentions of
programmers in all cases.

\bibliographystyle{acmtrans}

\end{document}